\documentclass[reprint,showpacs,amsmath,amssymb,xcolor=dvipsnames,prl,longbibliography,superscriptaddress,floatfix]{revtex4-1}
\pdfoutput=1

\usepackage[T1]{fontenc}
\usepackage[pdftex]{graphicx}% Include figure files
\usepackage{bm}% bold math
\usepackage[pdftex,colorlinks=true,citecolor=blue,urlcolor=blue,linkcolor=black]{hyperref}
\usepackage{braket}
\usepackage[dvipsnames]{xcolor}
\usepackage[version-1-compatibility,binary-units,abbreviations=true]{siunitx}
\usepackage{comment}

\usepackage{physics}
\usepackage{mathrsfs}
\usepackage{xspace}
\usepackage{hyperref}

\newcommand{\oR}{\omega}
\newcommand{\oZ}{\omega_\text{z}}
\newcommand{\com}{\text{com}}
\newcommand{\rel}{\text{rel}}

\definecolor{uc_blue}{HTML}{3891a6}
\definecolor{uc_red}{HTML}{c61a27}

\begin{document}

%%%%%%%%%%%%%%%%%%%%%%%%%%%%%%%%%%%%%%%%%%%%%%%%%%%%%%%%%%%%%%%%%%%%%%%%%%%%%%%%%
%						    	Title and Authors	    						%
%%%%%%%%%%%%%%%%%%%%%%%%%%%%%%%%%%%%%%%%%%%%%%%%%%%%%%%%%%%%%%%%%%%%%%%%%%%%%%%%%

\title{ Realization of a Laughlin State of Two Rapidly Rotating Fermions }

\author{Philipp~Lunt}
    \email{lunt@physi.uni-heidelberg.de}
	\affiliation{Physikalisches Institut der Universit\"at Heidelberg, Im Neuenheimer Feld 226, 69120 Heidelberg, Germany}
\author{Paul~Hill}
	\affiliation{Physikalisches Institut der Universit\"at Heidelberg, Im Neuenheimer Feld 226, 69120 Heidelberg, Germany}
\author{Johannes~Reiter}
	\affiliation{Physikalisches Institut der Universit\"at Heidelberg, Im Neuenheimer Feld 226, 69120 Heidelberg, Germany}
\author{Philipp~M.~Preiss}
	\affiliation{Max Planck Institute of Quantum Optics, Hans-Kopfermann-Str. 1, 85748 Garching, Germany}
	\affiliation{Munich Center for Quantum Science and Technology (MCQST), Schellingstr. 4, 80799 München, Germany } 
\author{Maciej~Ga\l ka}
	\affiliation{Physikalisches Institut der Universit\"at Heidelberg, Im Neuenheimer Feld 226, 69120 Heidelberg, Germany}
\author{Selim~Jochim}
	\affiliation{Physikalisches Institut der Universit\"at Heidelberg, Im Neuenheimer Feld 226, 69120 Heidelberg, Germany}
\date{\today  }

%%%%%%%%%%%%%%%%%%%%%%%%%%%%%%%%%%%%%%%%%%%%%%%%%%%%%%%%%%%%%%%%%%%%%%%%%%%%%%%%%
%						        	Abstract        	   						%
%%%%%%%%%%%%%%%%%%%%%%%%%%%%%%%%%%%%%%%%%%%%%%%%%%%%%%%%%%%%%%%%%%%%%%%%%%%%%%%%%

\begin{abstract}
We realize a Laughlin state of two rapidly rotating fermionic atoms in an optical tweezer. By utilizing a single atom and spin resolved imaging technique, we sample the Laughlin wavefunction thereby revealing its distinctive features, including a vortex distribution in the relative motion, correlations in the particles' relative angle, and suppression of the interparticle interactions. Our work lays the foundation for atom-by-atom assembly of fractional quantum Hall states in rotating atomic gases. 
\end{abstract}
\maketitle

%%%%%%%%%%%%%%%%%%%%%%%%%%%%%%%%%%%%%%%%%%%%%%%%%%%%%%%%%%%%%%%%%%%%%%%%%%%%%%%%%
%						        	Main Text         	   						%
%%%%%%%%%%%%%%%%%%%%%%%%%%%%%%%%%%%%%%%%%%%%%%%%%%%%%%%%%%%%%%%%%%%%%%%%%%%%%%%%%

Neutral particles in a rotating frame mimic the motion of charged particles subjected to a magnetic field~\cite{Dalibard_2016Introduction}. The Coriolis force takes on the role of the Lorentz force which dictates free particles to move on cyclotron orbits. Within a quantum mechanical framework, this results in quantized energy levels, referred to as Landau levels. They are infinitely degenerate in the case of translational invariance, separated by the large cyclotron frequency.  
The properties of such a system are classified by the filling factor $\nu$, which is defined as the ratio of the particle number and the number of states per Landau level~\cite{Giuliani}. For integer values, the Landau levels are completely occupied resulting in the integer quantum Hall effect~\cite{Klitzing_PhysRevLett.45.494}, while for $\nu < 1$ the lowest Landau level (LLL) is partially filled leading to the emergence of strongly correlated phases including fractional quantum Hall states~\cite{Tsui_PhysRevLett.48.1559, Wen}. The fractional filling factors $\nu = 1/m$, with an integer $m$, are qualitatively described by Laughlin's wavefunction~\cite{Laughlin_PhysRevLett.50.1395}.

\begin{figure}[t]
    \centering
	\includegraphics{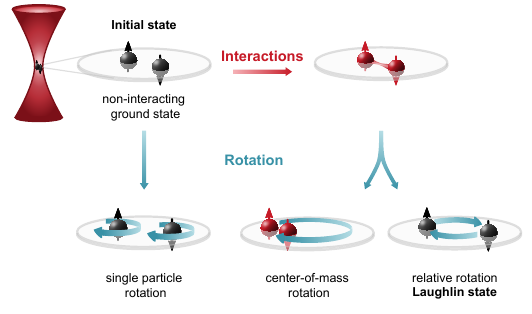}
    \caption{
    Conceptual path to a Laughlin state with two rapidly rotating fermions.
    Two spinful non-interacting fermions (black particles) are prepared in the ground state of a radially symmetric optical tweezer. To couple the particles to states with angular momentum, we tune the trap to an elliptical shape that rotates. 
    Rotating the non-interacting particles leads to independent, single-particle rotation. Interactions (indicated by red color) break the symmetry between the particles' center-of-mass and relative motion since the interactions only couple to the relative part of the wavefunction. Hence, states with angular momentum only in the center-of-mass degree of freedom have different energy than states with the same angular momentum in the relative motion and thus both can be selectively addressed with different rotation frequencies. States with angular momentum only in the relative degree of freedom are Laughlin states. These states possess a node in their relative wavefunction which, in our case of contact interactions, renders them non-interacting.
    }
    \label{fig:setup}
\end{figure}

Rotating ultracold atomic gases~\cite{Cooper_2008, Fetter_RevModPhys.81.647} is one approach among other techniques~\cite{Lin_2011, Chalopin_2020, Mancini_2015, Zhou_2023, Struck_PhysRevLett.108.225304, Aidelsburger_2014, Jotzu_2014, Kennedy_2015, Tai_2017, Asteria_2019} to study quantum many-body physics in magnetic fields.
In the slow-rotation limit of large filling factors $\nu \gg 1 $ quantized flux vortices are formed~\cite{Madison_PhysRevLett.84.806} and arrange in a triangular Abrikosov lattice~\cite{Abo-Shaeer_doi:10.1126/science.1060182, Zwierlein_2005}.
Reaching filling factors \textcolor{black}{ ${\nu \sim 100}$} has been achieved with ultracold Bose gases, signaled by the softening of the Abrikosov lattice~\cite{Schweickhardt_PhysRevLett.92.040404, Bretin_PhysRevLett.92.050403} and more recently by distilling a single Landau gauge wavefunction in the LLL~\cite{Fletcher_2021, Mukherjee_2022, yao2023observation}. 
In the limit of rapid rotation ${\nu \lesssim 1}$ strongly correlated phases are predicted analogous to phases occurring in the fractional quantum Hall effect~\cite{Cooper_2008, Cooper_QuantumPhases_PhysRevLett.87.120405, Popp_PhysRevA.70.053612, Regnault_PhysRevLett.91.030402}, where first \textcolor{black}{attempts have been pursued} in rotating atomic clusters of interacting bosons~\cite{gemelke2010rotating}; \textcolor{black}{in different systems,} Laughlin states with two photons~\cite{Clark_2020} and with two bosonic atoms in a driven optical lattice~\cite{Leonard_2023} \textcolor{black}{have been realized lately}. 
Ultracold Fermi gases at filling factors $\nu \lesssim 1$ have thus far been unexplored experimentally due to challenges to transfer fermions to the LLL via rotation of the potential, given the Pauli exclusion principle.

\begin{figure*}
    \centering
	\includegraphics{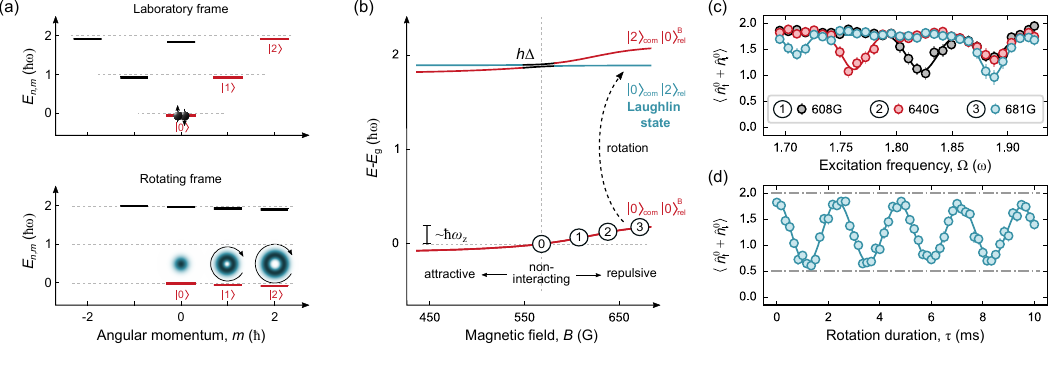}
    \caption{
    Preparation of the Laughlin state.
    \textcolor{black}{ (a) Two non-interacting fermions are prepared in the ground state of the optical tweezer (top). In-plane it forms an approximate harmonic oscillator. States $|m\rangle$ (red) with maximal angular momentum along the axial direction $L_\text{z} = m\hbar$ in their respective shell $n$ ($n=m$) form the lowest Landau level in the rotating frame with $\oR$, i.e. the deconfinement limit (bottom). }
    (b) Energy spectrum in the laboratory frame in the $L_\text{z}=0\hbar, 2\hbar$ angular momentum manifold of two interacting fermions in a cigar-shaped harmonic trap including an anharmonicity $\Delta$. 
    Adiabatic tuning of the magnetic field (\textcircled{0} $\rightarrow$ \textcircled{3}) leads to an energy shift $E_\text{int} $  with respect to the non-interacting ground state energy $E_\text{g}$ (horizontal dotted line); the maximum $E_\text{int} $ is on the order of the axial trap frequency $\omega _\text{z}$. 
    In the $L_\text{z}=2\hbar$ manifold in the limit $E_\text{int} \gg \Delta$, the two states correspond to $|2\rangle _\text{com} |0\rangle  _\text{rel}$ and $|0\rangle _\text{com} |2\rangle _\text{rel}$ in the center-of-mass and relative basis, with the latter being the Laughlin state $|\psi _\text{1/2}\rangle $.
    %,, i.e. when the center-of-mass and relative motion decouple. 
    (c) Excitation spectrum at various magnetic fields measured by counting the remaining atoms in the single-particle ground state $\langle\hat{n}^{0}_{\uparrow}+\hat{n}^{0}_{\downarrow}\rangle $ after applying the rotating perturbation for $\tau = \SI{350}{\mu s}$. For increasingly repulsive interacting atoms, the frequency of the relative excitation $\Omega _\text{rel} < 1.9 \oR$ decreases, while the frequency of center-of-mass excitation at $\Omega _\text{com} \sim 1.9 \oR$ stays approximately constant.
    (d) Rabi oscillations at \SI{680}{G} on the resonance $\Omega _\text{rel}  \approx \SI{1.7}{} \oR$ driven with a Rabi rate $\Omega _\text{rabi} / 2 \pi \approx \SI{0.42}{kHz}$. According to eq.~\ref{eq:laughlin_single_particle} the Rabi oscillations reach \SI{2}{} and \SI{0.5}{} (dashed lines).
    The error bars in (c,d) represent the standard error of the mean and are smaller than the data points if not visible.
    }
    \label{fig:excitation_spectrum}
\end{figure*}

In this Letter, we realize the  $\nu=1/2$ Laughlin state of two rapidly rotating spinful fermions in an optical tweezer. Our approach relies on the smooth rotation of the optical potential and precise control of the in-plane anisotropy to a level of $\SI{4e-4}{}$~\cite{SupMat}. On a conceptual level, the realization of the Laughlin state requires interactions between the particles and a strong magnetic field here engineered via rotation, illustrated in Fig.~1. In the absence of interactions, rotation of the particles leads to independent, single-particle rotation. In this scenario, the center-of-mass and relative motion of the particles are identical.
The interactions break the symmetry between the particles' center-of-mass and relative motion (defining the preferred basis) since they couple only to the relative part of the wavefunction. The interaction energy of a state with only center-of-mass angular momentum tunes differently compared to a state with relative angular momentum, allowing for selective spectroscopic addressing.

The Laughlin state for any particle number $N$ incorporates angular momentum in the relative motion between each and every particle, thereby suppressing the interaction energy. The spatial part of the Laughlin wavefunction is described by
\begin{equation}
    \psi _{1/m} (z_\text{1},...,z_\text{N}) = \prod _\text{i<j}   (z_\text{i} - z_\text{j})^ m e ^ { - \sum _\text{i=1}^{N} |z_\text{i} |^2 /2 },
    \label{eq:laughlin}
\end{equation}
where ${z_\text{j}=(x_\text{j}+i y_\text{j})/l_\text{HO}}$ labels the complex coordinate of the $j$th particle in the radial plane in units of the harmonic oscillator length $l_\text{HO}$, which defines the natural length scale of our system,  $m$ is the angular momentum in units of $\hbar$ incorporated in the relative motion of the particles $(z_\text{i}-z_\text{j} )$, and the filling factor relates via $\nu = 1 / m$.  In our two-particle case of a spin singlet, the spatial wavefunction is required to be symmetric, thereby restricting the exponent $m$ to even numbers.
In the many-body limit, our system connects to spinful fractional quantum Hall states described by the Halperin wavefunction, which is a generalization of the Laughlin wavefunction containing a spin degree of freedom~\cite{Halperin_1983_HelvPhysActa}.

\begin{figure*}[t]
    \centering
	\includegraphics{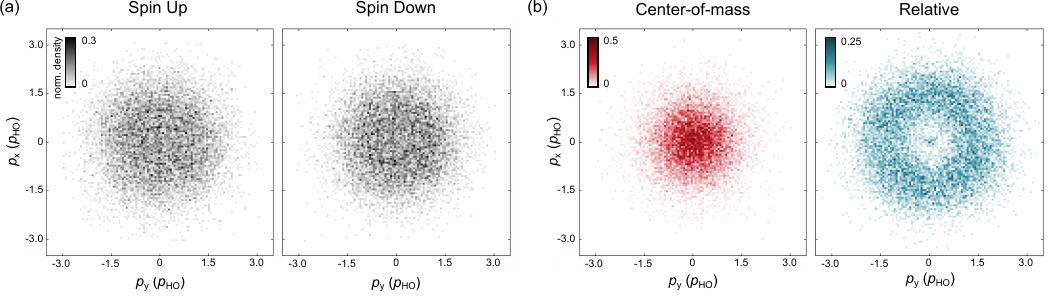}
    \caption{
    Laughlin wavefunction of two rapidly rotating fermions.
    (a) Normalized single-particle density of the spin up and spin down atom. Both densities flatten out for small momenta. 
    (b) Normalized density in center-of-mass coordinates $\textbf{p} _\text{com} = 1/\sqrt2 \left( \textbf{p} _\uparrow + \textbf{p} _\downarrow \right)$ and in relative coordinates $\textbf{p} _\text{rel} = 1/\sqrt2 \left( \textbf{p} _\uparrow - \textbf{p} _\downarrow \right)$. The center-of-mass motion has a Gaussian shape. The angular momentum is incorporated in the relative motion of the particles, resulting in an azimuthally symmetric vortex distribution of the density. The total angular momentum of $2\hbar$ determines the maximum of the vortex distribution at a radius of $\sqrt{2}p_\text{HO}$. Note that a small admixture $\sim \SI{1}{\%}$ of a Feshbach molecule is visible as a peak at zero relative momenta. 
    }
    \label{fig:laughlin_state}
\end{figure*}

\subsection{Preparation of the Laughlin state}
We start by preparing a non-interacting spin up and down fermion of $^6\text{Li}$ atoms in the ground state of a radially symmetric, tightly focused optical tweezer~\cite{Serwane_2011}. The tweezer forms a cigar-shaped harmonic trap with the radial $\oR / 2 \pi \approx \SI{56.1}{kHz}$ and axial $\oZ / 2 \pi  \approx \SI{7.9}{kHz}$ trap frequency, determining $l_\text{HO}=\sqrt{\hbar / m_\text{Li} \oR }$. 
Since the rotation couples only to the in-plane motion, we consider a two-dimensional (2D) potential in Fig.~2(a, top), which forms a rotationally symmetric 2D harmonic oscillator up to small anharmonic corrections. The energies $E_{n,m}$ are labeled by their shell number $n$ and their angular momentum $L_\text{z}=m\hbar$ along the axial direction. \textcolor{black}{When viewed in the reference frame rotating at $\oR$, the states remain eigenstates of the system, yet the energy rearranges such that the eigenstates form degenerate Landau levels,  shown in Fig.~\ref{fig:excitation_spectrum}(a, bottom). The states $|m\rangle$ (red states) with maximal angular momentum in each shell, i.e. $n=m$, belong to the LLL~\cite{SupMat}.}

To excite the atoms selectively to angular momentum eigenstates in the LLL, we interfere the tweezer with a Laguerre-Gaussian beam which deforms the originally radially symmetric potential and acts as a rotating perturbation $ {\propto (z^l e^{-i \Omega t } + h.c.)}$. It couples states that differ in angular momentum by $\hbar l $ and in energy by $\hbar \Omega $. Here, $2\pi l$ is the phase winding of the Laguerre-Gaussian beam, and $\Omega / l $ is the angular rotation frequency of the optical potential set by the angular frequency difference $\Omega $ between the tweezer and the perturbation beam. To realize the $\nu = 1/2$ Laughlin state, we introduce a total angular momentum of $2\hbar$ using the Laguerre-Gaussian beam with $l=2$~\cite{Lunt2024_pra}.

Our approach to reach the two-particle Laughlin state $\ket{\psi _\text{1/2}}$ is illustrated in Fig.~2(b). We outline the relevant energy levels of two contact-interacting fermions trapped in a cigar-shaped potential\textcolor{black}{, viewed in the laboratory frame. We express the eigenstates in the center-of-mass and relative coordinates}, as the interactions depend only on the relative motion and the center-of-mass and relative degrees of freedom decouple in a harmonic potential. The decoupling also holds in the presence of the weak anharmonicity $\Delta \approx \SI{1.4}{kHz}$ in the limit of large interaction energy $E_\text{int} / h \gg  \Delta$.

The normalized spatial wavefunction of the  two-particle Laughlin state $\ket{\psi _\text{1/2}}$ can be expressed using harmonic oscillator eigenstates
\begin{subequations}
    \begin{align}
        \ket{\psi _\text{1/2}} &= |0\rangle _\text{com} |2\rangle _\text{rel} \label{eq:laughlin_com_rel} \\
        &=  ( | 0 \rangle _\uparrow | 2 \rangle _\downarrow +  | 2 \rangle _\uparrow | 0 \rangle _\downarrow ) / 2 - | 1 \rangle _\uparrow | 1 \rangle _\downarrow  /\sqrt{2} ,\label{eq:laughlin_single_particle}
    \end{align}
\end{subequations}
in the center-of-mass and relative basis (eq. \ref{eq:laughlin_com_rel}) and the single-particle basis (eq. \ref{eq:laughlin_single_particle}). The Laughlin state  $\ket{\psi _\text{1/2}}$ remains in the center-of-mass ground state, while carrying $2\hbar$ angular momentum in the relative degree of freedom. 
In the single-particle basis, $\ket{\psi _\text{1/2}}$ is a superposition of states in the LLL, where either one spin state carries $2\hbar$ angular momentum leaving the other spin state in the ground state or each spin state carries $1\hbar$ angular momentum. Note, that we directly relate each particle to a spin value via the subscripts $\uparrow, \downarrow$~\footnote{\textcolor{black}{We prepare a spin singlet state $\ket{\Psi} = 1/\sqrt{2}(\ket{\uparrow \downarrow} - \ket{\downarrow \uparrow}) \otimes \ket{\psi _{1/2}}$ in each experimental realization. Since we do not couple to the spin degree of freedom and our fluorescence imaging measures in the $\{\ket{\downarrow},\ket{\uparrow}\}$ basis, the wavefunction in Eq.~(\ref{eq:laughlin_single_particle}) describes the spatial wavefunction after relating each particle to its spin value}}.

\begin{figure*}[t]
    \centering
	\includegraphics{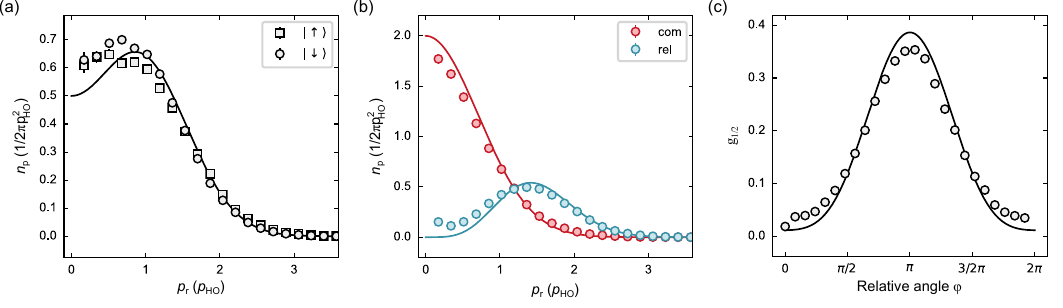}
    \caption{
    Properties of the Laughlin state.
    (a) Radial densities of the spin up and spin down fermion. The solid line is a mixture of the states $|0\rangle, |1\rangle, |2\rangle$, according to eq. \ref{eq:laughlin_single_particle}. 
    (b) The Laughlin state remains in the ground state $|0\rangle _\text{com}$ in center-of-mass coordinates, and occupies the $|2\rangle _\text{rel}$ state in relative coordinates. The relative density is not zero at small momenta due to the admixture of a Feshbach molecule. The solid lines are calculated according to eq.~\ref{eq:laughlin_com_rel}.
    (c) Normalized histogram of relative angle correlations between the spin up and spin down fermion.  The solid line is the theoretical angle correlation function $  {g _\text{1/2} (\varphi ) }$, see text. 
    Error bars of the 95\% confidence interval, determined using a bootstrapping technique, are smaller than the point size if not visible.
    }
    \label{fig:correlations}
\end{figure*}

We spectroscopically characterize the system by measuring the single-particle occupation number in the ground state $\langle\hat{n}^{0}_{\uparrow}+\hat{n}^{0}_{\downarrow}\rangle $ after applying the rotating perturbation for $\tau = \SI{350}{\mu s} $ at different interactions strengths, shown in Fig.~2(c). 
For those repulsive interactions, we observe two resonances which correspond to the center-of-mass $|2\rangle _\text{com} |0\rangle ^\text{B} _\text{rel}$ and relative rotation $|0\rangle _\text{com} |2\rangle _\text{rel}$ (for $E_\text{int}/h \gg\Delta$). Here, $|0\rangle ^\text{B} _\text{rel}$ labels the ground state in the relative motion which changes with the interaction strength tuned by the magnetic field. 
As the interaction energy depends only on the relative motion, the resonance frequency of the center-of-mass excitation tunes similarly to the ground state energy and thus stays approximately constant at $\Omega _\text{com} \sim 1.9 \oR$, down-shifted from $2\oR$ due to the anharmonicities. In contrast, the relative excitation $\Omega _\text{rel} < 1.9 \oR$ shifts to lower frequencies for larger repulsive interaction strengths. Modeling the energy levels in the trap allows us to subtract the interaction-dependent ground state energy, showing the suppression of inter-particle interactions in the $|0\rangle _\text{com} |2\rangle _\text{rel}$ state \textcolor{black}{(see also Fig.~S5 in \cite{SupMat})}.

By ramping the magnetic field to \SI{680}{G}, we reach the limit $E_\text{int} /h  \approx \SI{9.9}{kHz} \gg \Delta$. In Fig.~2(d), we show Rabi oscillations with a Rabi rate $\Omega _\text{rabi}/2\pi\approx \SI{0.42}{kHz}$ on the resonance $\Omega \approx \SI{1.7}{}\omega $ between the repulsively interacting ground state and the Laughlin state, to which we transfer via a $\pi-$pulse. Since we measure $\langle\hat{n}^{0}_{\uparrow}+\hat{n}^{0}_{\downarrow}\rangle $ we expect the minimum of the Rabi oscillations to reach $0.5$, following eq.~\ref{eq:laughlin_single_particle}. An upper bound of the preparation fidelity is inferred through the single-particle ground state occupation after half a Rabi cycle $\mathcal{F} _\text{prep} = \SI{96(2)}{\%} $, also accounting for the preparation fidelity of two atoms in the ground state.

Furthermore, we measure the lifetime of the Laughlin state by utilizing Ramsey spectroscopy which yields $\tau _\text{coh} = \SI{191(21)}{ms}$~\cite{SupMat}. Remarkably, this corresponds to \SI{21400}{} coherent rotations of the particles, demonstrating that the Laughlin state is insensitive to environmental noise sources.

\subsection{Observation of the Laughlin wavefunction}
We measure the density of the Laughlin wavefunction in momentum space after a time-of-flight expansion of ${t _\text{tof} = \SI{1.78}{ms}}$ \textcolor{black}{from approximately  $\SI{11000}{}$ experimental realizations} using a spin and atom resolved fluorescence imaging technique~\cite{Bergschneider_PhysRevA.97.063613}. 
Since the Laughlin state consists of states in the LLL and suppresses interactions the expansion corresponds to a magnification of the initial wavefunction~\cite{Cooper_PhysRevA.68.035601}.
We express all momenta in units of the harmonic oscillator momentum $p _\text{HO} = \sqrt{\hbar m \oR } $.

In Fig.~3(a), we show the normalized density in the single particle basis of the spin up and spin down fermion. In that basis, neither of the spin states exhibits a vortex distribution. Instead, the density has a flattened profile. Note that the density distribution of the spin up state is larger stemming from off-resonant scattering during the prior imaging of the spin down state~\cite{SupMat}.

In order to reveal the key signatures of the Laughlin wavefunction we transform to the center-of-mass and relative basis. In each experimental realization, we measure the momentum ${\textbf{p}=(p_\text{x},p_\text{y})}$ of the spin up and spin down fermion, denoted as $\textbf{p}_\uparrow$ and $\textbf{p}_\downarrow$, respectively. This allows us to calculate the center-of-mass $\textbf{p} _\text{com} = 1/\sqrt2 \left( \textbf{p} _\uparrow + \textbf{p} _\downarrow \right)$ and relative $\textbf{p} _\text{rel} = 1/\sqrt2 \left( \textbf{p} _\uparrow - \textbf{p} _\downarrow \right)$ coordinates in a single snapshot of the wavefunction. We determine the normalized density in the center-of-mass and relative basis in Fig.~3(b), revealing the striking features of the Laughlin wavefunction.

While in center-of-mass coordinates the density distribution of the Laughlin state has a Gaussian shape, it shows a rotationally symmetric vortex distribution in relative coordinates owing to the angular momentum incorporated in the relative motion of the particles. The amount of angular momentum $m=2\hbar$ determines the size of the vortex, with the peak density occurring at $\sqrt{2}p_\text{HO}$.
A small peak in the density at zero momenta is visible, stemming from anharmonic coupling to molecular states with center-of-mass excitations during the ramp of the magnetic field~\cite{Sala_PhysRevLett.110.203202}.

In Fig.~4, we compare our experimental data to theoretical predictions based on the Laughlin wavefunction $\psi _\text{1/2}$. In Fig.~4(a,b), we show the azimuthally averaged density distributions $n _\text{p}$ as a function of the radial momentum ${p_\text{r} = \sqrt{p_\text{x} ^2 + p_\text{y}^2} }$, in both the single-particle basis and center-of-mass and relative basis, respectively. In the single-particle basis, the Laughlin state is a superposition of states in the LLL, according to eq. \ref{eq:laughlin_single_particle}. In the center-of-mass and relative basis, the Laughlin state remains in the ground state $|0\rangle _\text{com}$ (red) and occupies the $|2\rangle _\text{rel} $ state (blue), respectively, according to eq. \ref{eq:laughlin_com_rel}. The measured densities qualitatively agree with the theoretical predictions (solid lines) without free parameters.
Additionally, we gain information about the phase of the wavefunction by letting the system evolve in a slightly anisotropic potential; the time evolution is consistent with the expected phase winding of $4 \pi$ corresponding to an angular momentum of $2\hbar$ (see Fig.~\ref{som:fig:ramsey_spectrum} in~\cite{SupMat}).

Furthermore, we extract the relative angle correlations between the fermions, similar to \cite{Clark_2020}. For that, we subtract the azimuthal angle of the spin up and down atom determined in each experimental realization and calculate the normalized angle distribution, shown in Fig.~4(c). The solid line represents the theoretical angle correlations of the ${\nu=1/2}$ Laughlin wavefunction 
${g _\text{1/2} (\varphi) = (6 - 3 \pi \cos \varphi  + 4\cos ^2 \varphi )/16\pi  }$, without free parameters.
In general, the Laughlin wavefunction $\psi _{1/m}$ exhibits a distinctive peak at a relative angle ${\varphi = \pi }$ between two particles, which becomes sharper with increasing angular momentum $m$~\cite{SupMat}. 
In the context of our zero-range contact interactions, this observation demonstrates that while the Laughlin state is non-interacting due to the node in the particles' relative wavefunction, it is strongly correlated in the motional degree of freedom.

To connect to the electronic fractional quantum Hall effect, we emphasize that the Laughlin state, which is an excited eigenstate of the system in the laboratory frame, becomes the ground state in the reference frame rotating at the deconfinement limit. In order to further relate the two-particle Laughlin wavefunction to the many-body limit, we consider the density  ${ \Tilde{n}_\text{p} = n_{\text{p}\uparrow}+n_{\text{p}\downarrow}}$ at its center. Since the two-particle Laughlin wavefunction is an eigenstate of the harmonic oscillator, the momentum space $\Tilde{n}_\text{p}$ and real space $\Tilde{n}_\text{r}$ densities are the same  ${2 \pi \Tilde{n}_\text{p} (p_r \rightarrow 0) p^2_\text{B}=2 \pi \Tilde{n}_r (r \rightarrow 0) l ^2_\text{B}} $ when expressed in units of the magnetic momentum $p_\text{B}$ and magnetic length $l_\text{B}$, which in our analogous system are ${p_\text{B}=\sqrt{2} p_\text{HO}}$ and ${l_\text{B}=l_\text{HO}/\sqrt{2}}$. The measured value $2 \pi \Tilde{n}_\text{p} (p_r \rightarrow 0) p^2_\text{B} \approx \SI{0.6}{}$ is close to the expected value of 1/2, marking the precursor of the incompressible bulk plateau of the fractional quantum Hall droplet, extending up to a radius ${2\sqrt{(N-1)} l_\text{B}}$~\cite{Cooper_2008}. In this region, the density increases before it falls off to zero on a scale of $l_\text{B}$, indicating the onset of the compressibility of the edge in the many-body limit~\cite{Wen}.

\subsection{Conclusion and Outlook}
We directly observe microscopic correlations of the ${\nu = 1/2}$ Laughlin wavefunction demonstrating its non-interacting, yet strongly correlated nature due to the incorporation of angular momentum in the relative motion of the contact-interacting fermions. Our platform opens up a new pathway to explore microscopic details of spinful fractional quantum Hall states with ultracold atoms. Future work involves the scalability to larger particle numbers to study the emergence of topological phases of matter~\cite{palm2023growing, Letscher_PhysRevB.91.184302}, the exploration of quantum Hall ferromagnetism~\cite{Palm_2020, Yang_PhysRevLett.72.732}, or the investigation of topologically distinct quantum phase transitions in the BEC-BCS crossover~\cite{Yang_PhysRevLett.100.030404, Repellin_PhysRevB.96.161111, ho2016fusing}.

%%%%%%%%%%%%%%%%%%%%%%%%%%%%%%%%%%%%%%%%%%%%%%%%%%%%%%%%%%%%%%%%%%%%%%%%%%%%%%%%%
%					     Acknowledgements and Contributions   		   			%
%%%%%%%%%%%%%%%%%%%%%%%%%%%%%%%%%%%%%%%%%%%%%%%%%%%%%%%%%%%%%%%%%%%%%%%%%%%%%%%%%

\paragraph*{Acknowledgements} We gratefully acknowledge insightful discussions with Lauriane Chomaz, Fabian Grusdt and Nathan Goldman. 
\paragraph*{Funding}
This work has been supported by the Heidelberg Center for Quantum Dynamics, the DFG Collaborative Research Centre SFB 1225 (ISOQUANT), Germany’s Excellence Strategy EXC2181/1-390900948 (Heidelberg Excellence Cluster STRUCTURES) and the European Union’s Horizon 2020 research and innovation program under grant agreements No.~817482 (PASQuanS),  No.~725636 (ERC QuStA) and No. 948240 (ERC UniRand). This work has been partially financed by the Baden-Württemberg Stiftung. 
 
\paragraph*{Author Contributions}
P.L. led the experimental work, with significant contributions from P.H.. 
P.L., P.H., J.R., and M.G. conducted the experiment. 
P.H. performed guiding theoretical studies. 
P.M.P. conceived the original ideas. 
P.M.P., M.G., and S.J. supervised the project.
P.L., M.G., and P.H. wrote the manuscript with input from all authors.
All authors contributed to the discussion of the results.

%%%%%%%%%%%%%%%%%%%%%%%%%%%%%%%%%%%%%%%%%%%%%%%%%%%%%%%%%%%%%%%%%%%%%%%%%%%%%%%%%%					     Bibliography                          		   			%
%%%%%%%%%%%%%%%%%%%%%%%%%%%%%%%%%%%%%%%%%%%%%%%%%%%%%%%%%%%%%%%%%%%%%%%%%%%%%%%%%


%%%%%%%%%%%%%%%%%%%%%%%%%%%%%%%%%%%%%%%%%%%%%%%%%%%%%%%%%%%%%%%%%%%%%%%%%%%%%%%%%
%								Supplemental Material							%
%%%%%%%%%%%%%%%%%%%%%%%%%%%%%%%%%%%%%%%%%%%%%%%%%%%%%%%%%%%%%%%%%%%%%%%%%%%%%%%%%

\setcounter{figure}{0}
\setcounter{equation}{0}
\renewcommand\theequation{S\arabic{equation}} 
\renewcommand\thefigure{S\arabic{figure}} 
\cleardoublepage
\newpage
\section*{Supplemental Material}

\paragraph*{\textbf{\textcolor{black}{Rotating frame of reference }}}

The Hamiltonian of a single, neutral particle confined in a 2D harmonic potential with angular frequency $\oR$, and rotating in a reference frame with angular frequency $\Omega _\text{rot}$, equals 
\begin{align}
\begin{split}
H &= \frac{\bm{p}^2}{2m_\text{Li}} + \frac{m_\text{Li}}{2}\oR ^2 \bm{r}^2 - \Omega _\text{rot}L_\text{z} \\
&= \frac{(\bm{p} - m_\text{Li} \bm{\Omega _\text{rot}} \times \bm{r} )^2}{2m_\text{Li}} + \frac{m_\text{Li}}{2} (\oR ^2 - \Omega _\text{rot} ^2) \bm{r}^2, 
\label{eq:single_particle_hamiltonian}
\end{split}
\end{align}
where $L_\text{z}$ is the angular momentum along the axial $z$ direction resulting from the transfer to the rotating frame; in the deconfinement limit ($\Omega _\text{rot}/\oR=1$) the Hamiltonian is equivalent to the Hamiltonian of a particle with charge $q$ moving in a magnetic field $B$.
Here, the rotation plays the role of a magnetic field $q B =2 m_\text{Li} \Omega _\text{rot}$.
The eigenspectrum can be calculated for any $\Omega _\text{rot}$ and the eigenenergies take the form~\cite{Cooper_2008}
\begin{align}
E_\text{n,m} / \hbar \oR =  (n + 1) -m  \Omega _\text{rot} / \oR,
\end{align}
where $n$ is the shell quantum number and $m$ is the angular momentum quantum number. The corresponding eigenfunctions in polar coordinates  are
\begin{align}
\varphi _{n,m} (r, \phi ) = \sqrt{ \frac{k!}{\pi (k+|m|)!}  }   r^{|m|} e^{im \phi } e^{ -r^2/2 }  L_k ^{|m|}\left( r^2  \right) ,
\label{eq:HO_eigenstates}
\end{align}
where $r$ is expressed in units of $l_\text{HO}$, and  $L_k ^m$ are the associated Laguerre polynomials of degree $k= 1/2 (n - |m|)$ and order $|m|$, with $n \geq 0 $ and  $m=-n,-n+2, ....,n-2, n$.

\begin{figure}
    \centering
	\includegraphics{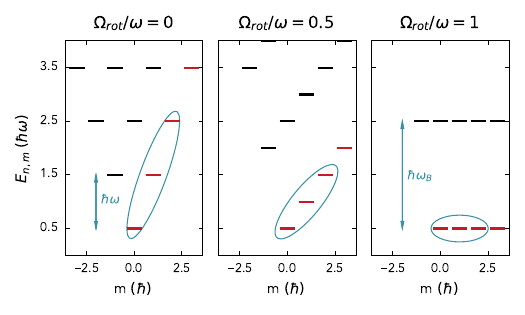}
    \caption{
    Energy levels of the 2D harmonic oscillator in the rotating frame.  
    (left) Stationary case $\Omega _\text{rot}/ \oR= \SI{0}{}$.
    (center) Rotating with half the trap frequency $\Omega _\text{rot}/ \oR= \SI{0.5}{}$. 
    (right) In the deconfinement limit $\Omega _\text{rot}/ \oR= \SI{1}{}$, the energy levels are highly degenerate and spaced by $\hbar \omega _\text{B} = 2 \hbar \oR$, analogous to the Landau levels of electrons subjected to a strong magnetic field. The circle highlights energy levels occupied by the $\nu=1/2$ two-particle Laughlin state
    \label{som:fig:energy_levels_rotation}
    }
\end{figure}

In Fig.~\ref{som:fig:energy_levels_rotation}, we show the energy levels for three different $\Omega _\text{rot}$. In the laboratory frame ${\Omega _\text{rot} / \oR = 0}$, the energy levels are spaced by the trap frequency $\oR$, where the $n$th energy level contains $n+1$ degenerate levels. This degeneracy is lifted with increasing rotation frequency (see central column ${\Omega _\text{rot} / \oR = \SI{0.5}{}}$). In the deconfinement limit ${\Omega _\text{rot}/ \oR = 1}$, the centrifugal force cancels the trapping frequency leading to highly degenerate energy levels which are spaced by the cyclotron frequency $\omega _\text{B} = 2 \oR$, analogous to the Landau levels of an electron subjected to a strong magnetic field. From the definition of the cyclotron frequency follows that in the deconfinement limit the analogous magnetic length is $l_\text{B} = \sqrt{\hbar / m_\text{Li} \omega _\text{B} } = l_\text{HO} / \sqrt{2}$ and equivalently the magnetic momentum $p_\text{B} = \sqrt{\hbar m_\text{Li} \omega _\text{B} }= \sqrt{2} p_\text{HO}$.
Note that the eigenfunctions $\varphi _{n,m} (r, \phi )$ do not depend on $\Omega _\text{rot}$, and thus the eigenstates of the system remain the same at all rotation frequencies, and in particular the eigenstes of the system are the same both in the laboratory frame and in the deconfinement limit.

The properties of the many-body system in the deconfinement limit, at which the degenerate Landau levels form, are defined by the filling factor~\cite{Cooper_2008}
\begin{align}
    \nu = 2 \pi n_\text{2D} l_\text{B}^2,
\end{align}
where $n_\text{2D}$ is the 2D density. Essentially, it describes the number of particles per available states in a Landau level. For filling factors $\nu < 1$, interactions start to play the dominant role, which leads to the formation of fractional quantum Hall liquids, among other phenomena. 
The class of $\nu = 1/m$ filling factors, where $m$ is an integer, are famously described by Laughlin's wavefunction, as defined in eq.~\ref{eq:laughlin} in the main text. Here, $m$ is even for spatially symmetric wavefunctions, such as spinless bosons or two spinful fermions in a spin singlet state, or $m$ is odd for spinless fermions.  
In the rotating frame, the $\nu=1/2$ Laughlin state consisting of $N$ bosonic particles is the exact ground state for zero-range, repulsive contact interactions at a total angular momentum ${M=N(N-1)}$~\cite{Wilkin_PhysRevLett.80.2265}.

In the case of two particles, the total angular momentum is $M=2$, leading to the occupation of the $m/\hbar=0,1,2$ angular momentum states (highlighted by a blue circle in Fig.~\ref{som:fig:energy_levels_rotation}). The $\nu=1/2$ Laughlin wavefunction for $N=2$ particles with spin up $\uparrow$ and spin down $\downarrow$, expressed in the complex coordinates $z=re^{i\phi}$, takes the form 
\begin{align}
\psi _{1/2} (z_\uparrow,z_\downarrow) &= (z_\uparrow - z_\downarrow )^2 e^{-(|z_\uparrow|^2 + |z_\downarrow|^2)/2 } \\
&= (z_\uparrow ^2 + z_\downarrow ^2 - 2 z_\uparrow z_\downarrow) e^{-(|z_\uparrow|^2 + |z_\downarrow|^2)/2 } \\
&\sim (\ket{2}_\uparrow \ket{0}_\downarrow + \ket{0}_\uparrow \ket{2}_\downarrow ) / 2-  \ket{1}_\uparrow \ket{1}_\downarrow / \sqrt{2}, 
\end{align}
where in the last line we use the fact that in the LLL (i.e. $n=m$) the eigenstates $\ket{m}$ scale with $\varphi _{m,m}\propto z^m$ (see eq.~\ref{eq:HO_eigenstates}); this wavefunction corresponds to eq.~\ref{eq:laughlin_single_particle} in the main text. Making a coordinate transformation to the center-of-mass and relative basis using $z_\com = 1/\sqrt{2} (z_\uparrow + z_\downarrow)$ and $z_\rel = 1/\sqrt{2}(z_\uparrow - z_\downarrow)$, results in
\begin{align}
\psi _{1/2} (z_\com, z_\rel ) &= \frac{1}{2}z_\rel ^2 e^{-(|z_\com|^2 + |z_\rel|^2)/2} \\
&\sim \ket{0}_\com \ket{2}_\rel ,
\end{align}
which corresponds to eq.~\ref{eq:laughlin_com_rel} in the main text.

\paragraph*{\textbf{Level and resonance spectrum of two interacting fermions in a Gaussian potential}}

\begin{figure*}
    \centering
	\includegraphics{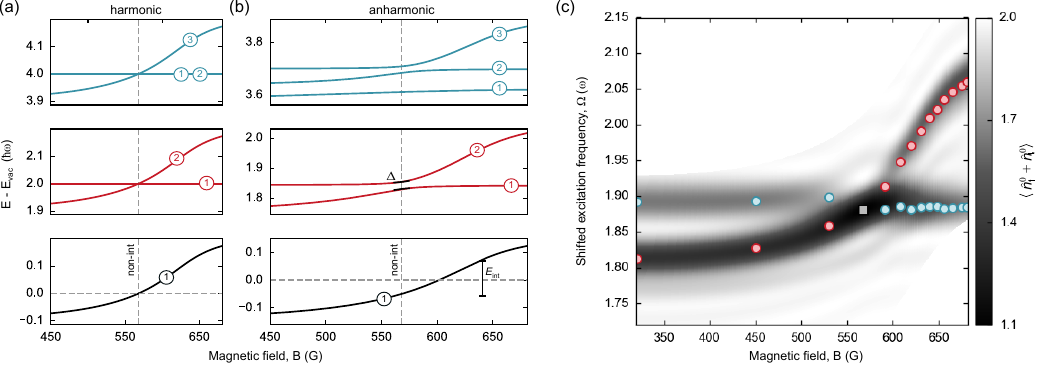}
    \caption{
    \textbf{Level and resonance spectrum of two interacting fermions in a Gaussian optical potential.}  
    (a) Subset of the energy spectrum of two interacting particles in a harmonic potential. Only the lowest lying states (with even exchange symmetry) of the angular momentum manifolds with $m=0$ (black, $\ket{\textcolor{black}{\textcircled{1}}}=\ket{0}_\text{com}\ket{0}_\text{rel}^{\text{B}}$), $m=2$ (red, $\ket{\textcolor{uc_red}{\textcircled{1}}}=\ket{0}_\text{com}\ket{2}_\text{rel}$, $\ket{\textcolor{uc_red}{\textcircled{2}}}=\ket{2}_\text{com}\ket{0}_\text{rel}^{\text{B}}$), and $m=4$ (blue, $\ket{\textcolor{uc_blue}{\textcircled{1}}}=\ket{2}_\text{com}\ket{2}_\text{rel}$, $\ket{\textcolor{uc_blue}{\textcircled{2}}}=\ket{0}_\text{com}\ket{4}_\text{rel}$, $\ket{\textcolor{uc_blue}{\textcircled{3}}}=\ket{4}_\text{com}\ket{0}_\text{rel}^{\text{B}}$) are shown. The molecular ground state of the system (and any of its center-of-mass excitations) is omitted. Energies are given relative to the harmonic two-particle zero-point energy $E_\text{vac}/\hbar\omega = 2 + \eta$.
    (b) The same spectrum incorporating the effect of a quartic, anharmonic perturbation stemming from the anharmonic shape of a Gaussian optical tweezer. The perturbation introduces avoided crossings in the spectrum around the non-interacting point (\SI{568}{G}) and lifts any degeneracy present in the spectrum.
    (c) Experimental and theoretical resonance spectrum of two interacting fermions in a Gaussian optical potential. Colored data points are the experimentally determined resonance frequencies extracted through atom loss spectroscopy (see main text Fig.~2c). In blue, the resonances we associate with the lowest relative excitation $|0\rangle _\text{com}|2\rangle _\text{rel}$ and in red, with the lowest center-of-mass excitation $|2\rangle _\text{com}|0\rangle _\text{rel}^\text{B}$, respectively, for magnetic fields at which $E_\text{int} \gg \Delta $.
    The numerical calculations determine the single-particle ground state occupation $\langle\hat{n}^{0}_{\uparrow}+\hat{n}^{0}_{\downarrow}\rangle $ after applying a rotating perturbation for a duration $\tau = \SI{350}{\mu s}$.
    The energy shift of the interaction dependent ground state in the anharmonic potential (see lower row in (b)) is subtracted from both the experimentally measured and numerically calculated excitation frequencies, showing that, within the model, the interactions are suppressed for the relative excitation.
    \label{som:fig:resonance_spectrum}
    }
\end{figure*}

Despite being analytically solvable the problem of two harmonically trapped particles interacting via a contact potential already gives rise to a rich, complicated level spectrum, strongly depending on the dimensionality of the problem \cite{Idziaszek_PhysRevA.74.022712}. Remarkably, the decoupling of the center-of-mass and relative degrees of freedom remains unbroken in the harmonic trap. The problem separates and only the relative motion is interacting,
\begin{equation}
\begin{split}
    H/\hbar\oR &= H_\text{com} + H_\text{rel}\\
    H_\text{com} &= \frac{1}{2}\Delta_\text{com} + \frac{1}{2}r_\text{com}^2 + \frac{1}{2}\eta^2 z_\text{ax}^2\\ 
    H_\text{rel} &= \frac{1}{2}\Delta_\text{rel} + \frac{1}{2}r_\text{rel}^2 + \frac{1}{2}\eta^2 z_\text{ax}^2 + g\delta^{(3)}(\bm{r}_\text{rel}), \label{eq:H2p}
\end{split}
\end{equation}
with $\bm{r}_\text{com, rel} = (\bm{r}_1\pm\bm{r}_2)/\sqrt{2}$, and lengths being expressed in harmonic oscillator units $l_\text{HO}$. To avoid confusion with the complex radial coordinate $z=x+iy=re^{i\varphi}$, we denote the axial coordinate by $z_\text{ax}$. Furthermore, $\eta = \omega_\text{z}/\oR$ defines the aspect ratio in terms of the axial and radial trap frequencies and the coupling constant $g$ can be tuned via the magnetic field. Due to the short-range nature of the interaction, relative harmonic oscillator states with nodes at the center suppress the interaction and remain eigenstates of the Hamiltonian. These are all states containing non-zero angular momentum or having odd axial parity. Thus, the complete spectrum may be written in terms of non-interacting harmonic oscillator states ${\ket{n, m; n_z}_\text{com}\otimes\ket{n', m'; n_z'}_\text{rel}}$ (${m' \neq 0}$ or ${n_z'}$ odd) and interacting states ${\ket{n,m;n_z}_\text{com}\otimes\ket{n_\text{int}}^{\text{B}}_\text{rel}}$. Here $n$ and $m$ label the radial shell number and the axial projection of angular momentum as in the main text, $n_z$ labels axial excitations, $n_\text{int}$ indexes the interacting states, and the dependence of the state and its energy on the magnetic field B is indicated by the superscript. To describe the motion of two spinful fermions in a spin-singlet state we require even exchange symmetry of the spatial wavefunction, which is satifisfied by excluding all odd relative angular momenta and axial excitations, i.e. we have $m', n_z'$ even.

In our experiment, we work with a cigar-shaped trap with an aspect ratio of $\eta\sim 0.14$. Hence, the spectrum is crowded with axial excitations. However, since our perturbation predominantly couples degrees of freedom in the radial plane, and we work with Rabi rates $\Omega_\text{rabi}\ll\omega_\text{z}$, we consider states in the axial ground state. 
The population of the Feshbach molecule is not significant during the preparation on the repulsive side of the Feshbach resonance and hence ignored including all its center-of-mass excitations. The atoms then only populate the second interacting state ($n_\text{int}=1)$, the repulsive ground state, which connects to the non-interacting ground state $\ket{n'=0, m'=0; n_z'=0}_\text{rel}$ at the magnetic field where the scattering length vanishes (\SI{568}{G}). From now on, if not otherwise stated, we will drop the $n$-, and $n_z$-label, as we will only be interested in the lowest lying states for given angular momentum. In particular, the repulsive ground state is denoted by $\ket{0}_\text{rel}^{\text{B}}$. In Fig.~\ref{som:fig:resonance_spectrum}(a) we show the spectrum of the lowest energy states in the three angular momentum manifolds $m=0$ (black, $\ket{0}_\text{com}\ket{0}_\text{rel}^{\text{B}}$), $m=2$ (red, $\ket{0}_\text{com}\ket{2}_\text{rel}$, $\ket{2}_\text{com}\ket{0}_\text{rel}^{\text{B}}$), and $m=4$ (blue, $\ket{2}_\text{com}\ket{2}_\text{rel}$, $\ket{0}_\text{com}\ket{4}_\text{rel}$, $\ket{4}_\text{com}\ket{0}_\text{rel}^{\text{B}}$), respectively. Note the even exchange symmetry of the wavefunction, excluding states with odd relative angular momentum.

At \SI{568}{G} the spectrum in Fig.~\ref{som:fig:resonance_spectrum}(a) exhibits prominent level crossings, reflecting the degeneracy that arises when placing multiple independent particles in a potential with equidistant energy levels. This degeneracy will be lifted in the presence of a generic (single-particle) perturbation, in our case an anharmonic correction to the harmonic potential. Degenerate perturbation theory then gives rise to the states
\begin{equation*}
\begin{split}
    \ket{\textcolor{uc_red}{\textcircled{1}}} &= \frac{1}{\sqrt{2}}\left(\ket{2}\ket{0} + \leftrightarrow\right) = \frac{1}{\sqrt{2}}\Big(\ket{2}_\com\ket{0}_\rel + \leftrightarrow\Big)\\
    \ket{\textcolor{uc_red}{\textcircled{2}}} &= \ket{1} \ket{1} = \frac{1}{\sqrt{2}}\Big(\ket{2}_\com\ket{0}_\rel - \leftrightarrow\Big)
\end{split}
\end{equation*}
in the $2\hbar$-manifold, and
\textcolor{MidnightBlue}{
\begin{equation*}
\begin{split}
    \ket{\textcolor{uc_blue}{\textcircled{1}}} &=\frac{1}{\sqrt{2}}\left(\ket{4}\ket{0} + \leftrightarrow\right) = \sqrt{\frac{1}{8}}\Big(\ket{4}_\rel\ket{0}_\com+\leftrightarrow\Big) \\&\qquad\qquad\qquad\qquad\qquad\qquad\qquad+ \frac{\sqrt{3}}{2}\ket{2}_\com\ket{2}_\rel\\
    \ket{\textcolor{uc_blue}{\textcircled{2}}} &=\frac{1}{\sqrt{2}}\left(\ket{3}\ket{1} + \leftrightarrow\right) = \frac{1}{\sqrt{2}}\Big(\ket{4}_\rel\ket{0}_\com-\leftrightarrow\Big)\\
    \ket{\textcolor{uc_blue}{\textcircled{3}}} &= \ket{2} \ket{2} = \sqrt{\frac{3}{8}}\Big(\ket{4}_\rel\ket{0}_\com+\leftrightarrow\Big) - \frac{1}{2}\ket{2}_\com\ket{2}_\rel\\
\end{split}
\end{equation*}
}
in the $4\hbar$ - manifold, respectively (c.f. also eq. \eqref{eq:non-int_single_particle}) and gap openings in the spectrum. Here, the states are expressed in the single-particle basis of harmonic oscillator orbitals on the left-hand side, and in the center-of-mass and relative motion basis on the right-hand side, respectively. Furthermore, the symbol $\leftrightarrow$ means that the expression to its left is repeated with interchanged quantum numbers. In Fig. ~\ref{som:fig:resonance_spectrum}(b), we plot the perturbed level spectrum in an anharmonic trap computed as explained below. As expected, we observe avoided crossings around \SI{568}{G}. Note, that the states will in general not be spaced equidistantly any longer, enabling Rabi transitions to the states $\ket{1} \ket{1}$, and $\ket{2} \ket{2}$ to be driven with the respective Laguerre-Gaussian modes. Away from the level crossing the states mostly recover the original harmonic forms, e.g. for $B\rightarrow$ \SI{680}{G} (repulsive) one finds $\ket{\textcolor{uc_red}{\textcircled{1}}}\rightarrow\ket{0}_\com\ket{2}_\rel$, $\ket{\textcolor{uc_red}{\textcircled{2}}}\rightarrow\ket{2}_\com\ket{0}^{\text{B}}_\rel$, $\ket{\textcolor{uc_blue}{\textcircled{3}}}\rightarrow\ket{4}_\com\ket{0}^{\text{B}}_\rel$, while $\ket{\textcolor{uc_blue}{\textcircled{1}}}$ and $\ket{\textcolor{uc_blue}{\textcircled{2}}}$ approach superpositions of $\ket{0}_\com\ket{4}_\rel$ and $\ket{2}_\com\ket{2}_\rel$ as both states don't tune with interactions.

In our case, the dominant perturbation in the optical tweezer is given by its anharmonic shape. To include this effect we use a two-dimensional Gaussian beam model for our optical tweezer, i.e. we neglect any anharmonicity present in the axial direction. The (2D-) Hamiltonian of a single particle reads
\begin{equation}
H_\text{sp} = -\frac{\hbar^2}{2m _\text{Li}}\Delta - \gamma\frac{P}{W^2} e^{-2r^2/W^2} \label{eq:H_gauss} \quad \text{(SI-units)},
\end{equation}
with $\gamma \approx 800 ~h\cdot\text{kHz µm}^2 /\text{mW}$. Radial excitation frequencies then only depend on two parameters, the tweezer power $P$ and its waist $W$. We fit the transition frequencies obtained from this Hamiltonian to experimentally observed frequencies by optimizing both parameters. This yields a waist of \SI{1.1}{\micro m}.

To incorporate the effect of the anharmonicity into the two-particle spectrum we include the first anharmonic term arising from the Gaussian potential in eq. \eqref{eq:H_gauss}, $-2\gamma Pr^4/W^6$. This gives rise to a quartic term $\frac{1}{2}\alpha r_\com^4$ ($\frac{1}{2}\alpha r_\rel^4$) in $H_\com$ ($H_\rel$), as well as
\begin{equation}
\begin{split}
    V_\text{c}/\hbar\omega &= 2\alpha r_\com^2 r_\rel^2 + \frac{\alpha}{2}z_\com^2 \cdot z_\rel^{*2} + h.c.\\
\end{split}
\end{equation}
coupling center-of-mass and relative motion, with $\alpha = -2\gamma ml_\text{HO}^6P/\hbar^2W^6$.

Starting from the solutions of the harmonic case, we first treat center-of-mass and relative motion separately. We perform perturbation theory to first order resulting in energy shifts $\Delta E$ linear in $\alpha$ of the harmonic states. For interacting states, these shifts will in general depend on the inter-particle interaction, due to the changing overlap with the quartic, anharmonic term. To obtain the full spectrum, we restrict the Hilbert space to the lowest energy states of the $0\hbar$, $2\hbar$, and $4\hbar$ angular momentum manifolds as above, and then perform a full diagonalization including the coupling term, to obtain the level spectrum of the two particles. Numerical results are plotted in Fig.~\ref{som:fig:resonance_spectrum}(b).

\begin{figure*}
    \centering
	\includegraphics{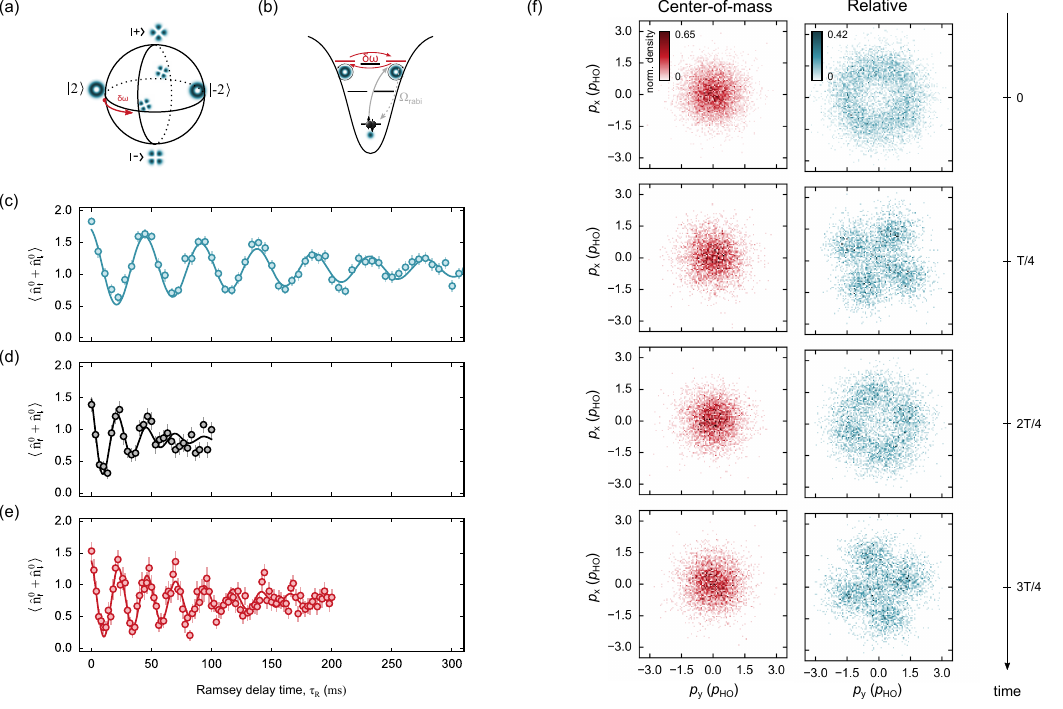}
    \caption{
    {Ramsey spectra.}  
    (a) Eigenstates in the anisotropic harmonic oscillator generically representing single-particle, center-of-mass, or relative motion states. In the presence of anisotropy the superposition states $|\pm \rangle = 1/\sqrt{2} (|2\rangle \pm |-2\rangle ) $ form the new eigenstates of the system with an energy difference given by the anisotropy $\delta \omega $. The initialized state $|2 \rangle $ is an equal superposition of the eigenstates $|+\rangle , |- \rangle $ and hence evolves over time $\tau$ on the equator of the Bloch sphere.
    (b) The anisotropy couples the clockwise $|2\rangle$ and counter-clockwise $|-2\rangle$ rotating states. We harness this effect to perform Ramsey spectroscopy on the $|2\rangle$ state by preparing it via a $\pi-$pulse. After a delay time $\tau _\text{R}$, we use a second $\pi -$pulse to de-excite the atoms to the ground state. Depending on the contribution of the $|-2\rangle$ state, the overlap with the ground state will oscillate.  
    (c) Ramsey spectrum of the Laughlin state. We measure an anisotropy of $\delta \omega / 2 \pi= \SI{21.4(1)}{Hz}$ at a trap frequency $\oR / 2 \pi \approx \SI{56.1}{kHz}$ yielding a relative anisotropy $\delta \oR / \oR = \SI{4e-4}{}$. 
    (d) Ramsey spectrum of two non-interacting particles at \SI{568}{G}. We measure an oscillation frequency of $\delta \oR ^\text{(non-int)} / 2 \pi = \SI{42(1)}{Hz}$. The coherence time is $\tau ^\text{(non-int)} _\text{coh} = \SI{33(8)}{ms}$.
    (e) Ramsey spectrum of the center-of-mass excitation $|2\rangle _\text{com} |0\rangle _\text{rel}^\text{B}$ at \SI{680}{G}. We measure an oscillation frequency of $\delta \oR ^\text{(com)} / 2 \pi = \SI{42.0(2)}{Hz}$. The coherence time is $\tau ^\text{(com)} _\text{coh} = \SI{87(12)}{ms}$.
    (f) Normalized density distribution in center-of-mass (red) and relative (blue) coordinates for different evolution times. The center-of-mass motion remains stationary in the ground state. The relative motion follows the evolution as explained in (a). The error bars in (c,d,e) correspond to the standard error of the mean and are smaller than the data point if not visible.
    \label{som:fig:ramsey_spectrum}
    }
\end{figure*}

To compare the theoretical model to our experimental data we compute the atom loss spectrum for a rotating LG$_{02}$-mode perturbation, which can be written as
\begin{equation}
\begin{split}
    V_\text{p} (t) &= \beta \left(z_1^2e^{-i\Omega t} + z_2^2e^{-i\Omega t} + h.c. \right)\\
            &= \beta \left(z_\com^2e^{-i\Omega t} + z_\rel^2e^{-i\Omega t} + h.c.\right).
\end{split}
\end{equation}
We compute the single-particle ground state occupation $\langle\hat{n}^{0}\rangle_t =\langle\hat{n}^{0}_{\uparrow}+\hat{n}^{0}_{\downarrow}\rangle_t $ after applying $V_\text{p}$ for a duration $t$. In a frame rotating with half the excitation frequency $\Omega$ the rotating perturbation appears constant and the initial state evolves according to
\begin{equation}
    \ket{\psi}_t = e^{-i\left[H+V_\text{p}(0)-\frac{1}{2}\Omega L_z\right]\frac{t}{\hbar}}\ket{0}_\text{com}\ket{0}_\text{rel}^{\text{B}}.
\end{equation}
Here, $\ket{0}_\text{com}\ket{0}_\text{rel}^{\text{B}}$ is the ground state of the system at some magnetic field B and the angular momentum operator $L_z$ appears due to the transformation to the rotating frame.

In Fig.~\ref{som:fig:resonance_spectrum}(c), we show numerical results of $\langle\hat{n}^{0}\rangle_t$ for varying excitation frequencies $\Omega$, and magnetic fields $B$ (different scattering lengths). Colored data points on top correspond to resonances extracted from the measured atom loss spectra; the resonance spectra at magnetic fields \SI{608}{G}, \SI{640}{G}, and \SI{681}{G} are shown in the main text in Fig.~2(c). From a single resonance at vanishing scattering length (\SI{568}{G}), we observe the emergence of two resonance branches when interactions are switched on. We associate these resonances with the lowest center-of-mass $\ket{2}_\com\ket{0}^{\text{B}}_\rel$ (red), and relative motion excitation $\ket{0}_\com\ket{2}_\rel$ (blue, Laughlin state) in the $2\hbar$ angular momentum manifold, for fields at which the interaction induced energy splitting between both levels becomes larger than the anharmonicity $\Delta$ and our Rabi rate $\Omega_\text{rabi}$. Note that the nature of the excited state changes drastically when approaching the non-interacting point. At zero interactions, where the degeneracy between center-of-mass and relative motion states is lifted by the anharmonicity, resulting in the avoided crossing (c.f. Fig.~\ref{som:fig:resonance_spectrum}b), one may expect to find two resonances, associated with the states $\ket{\textcolor{uc_red}{\textcircled{1}}}$, and $\ket{\textcolor{uc_red}{\textcircled{2}}}$. However, as our perturbation is symmetric between center-of-mass and relative motion only the $\ket{\textcolor{uc_red}{\textcircled{1}}}$ state can be excited leading to the disappearance of the upper resonance. Re-expressing the resonant state in terms of the single-particle basis, $\ket{\textcolor{uc_red}{\textcircled{1}}} \sim \ket{2}\ket{0} + \ket{0}\ket{2}$, we see that it represents an intermediate state appearing during the Rabi-oscillations between $\ket{0}\ket{0}$ and $\ket{2}\ket{2}$. Due to the underlying exchange symmetry of the particles, the intermediate and final state are still equidistantly spaced even in the presence of anharmonicity. In the center-of-mass and relative motion picture, we may then interpret the non-interacting Rabi oscillations as a process taking the atoms from the ground state to $\ket{\textcolor{uc_blue}{\textcircled{2}}} = \ket{2}\ket{2}$ in the $4\hbar$ angular momentum manifold, via the intermediate state $\ket{\textcolor{uc_red}{\textcircled{1}}}$. Note, that the perturbation in principle also couples $\ket{\textcolor{uc_red}{\textcircled{1}}}$ to $\ket{\textcolor{uc_blue}{\textcircled{1}}}$. However, this transition is suppressed due to the anharmonic level splitting for sufficiently small Rabi rates.

\paragraph*{\textbf{Time evolution of the Laughlin wavefunction}}

We measure the time evolution of the Laughlin wavefunction in a slightly anisotropic trap. Details on the strategy and experimental protocol, as well as the theoretical analysis are presented in the accompanying article \cite{Lunt2024_pra}, where we use the weak anisotropy to reveal the angular momentum carried by a single particle. The same scenario carries over to the two-particle case. Here, the system forms new eigenstates $\ket{\pm} = 1/\sqrt{2}\ket{0}_\com\left(\ket{2}_\rel \pm \ket{-2}_\rel\right)$ with an energy difference given by the anisotropy $\delta \omega $. The states $|\pm \rangle$  effectively form a two-level system,  which can be depicted on the Bloch sphere, shown in Fig.~\ref{som:fig:ramsey_spectrum}(a).

We perform Ramsey spectroscopy on the Laughlin state and observe coherent oscillations between the $|2\rangle _\text{rel}$ and $|-2\rangle  _\text{rel}$ states with a frequency given by the anisotropy $\delta \oR$. The conceptual protocol is sketched in Fig.~\ref{som:fig:ramsey_spectrum}(b). We use a $\pi-$pulse at \SI{680}{G} to inject $2\hbar$ quanta of angular momentum (gray solid arrow). Subsequently, we let the system evolve for a delay time $\tau _\text{R}$ (red arrows), after which we use a second $\pi-$pulse to de-excite the evolved state to the ground state (gray dashed arrow) and measure the single-particle occupation number in the ground state $\langle\hat{n}^{0}_{\uparrow}+\hat{n}^{0}_{\downarrow}\rangle $.

In Fig.~\ref{som:fig:ramsey_spectrum}(c,d,e), we show the measured Ramsey spectrum of the Laughlin state (blue), the non-interacting rotating fermions at \SI{568}{G} (black), and the center-of-mass excitation ${|2\rangle _\text{com} |0\rangle _\text{rel}^\text{B}}$ (red), respectively. 
Remarkably, the coherence time  of the Laughlin state ${\tau _\text{coh} = \SI{191(21)}{ms}}$ exceeds that of the non-interacting rotating atoms  ${\tau ^\text{(non-int)} _\text{coh} = \SI{33(8)}{ms}}$ and the ${|2\rangle _\text{com} |0\rangle _\text{rel}^\text{B}}$ state ${\tau ^\text{(com)} _\text{coh} = \SI{87(12)}{ms}}$.
We speculate that the extended coherence time %to the correlated motion of the particles in the Laughlin state (in contrast to $|\psi ^\text{(non-int)} \rangle$) 
can be attributed to the independence of the Laughlin state energy on the external magnetic field (in contrast to both $|\psi ^\text{(non-int)} \rangle$ and ${|2\rangle _\text{com} |0\rangle _\text{rel}^\text{B}}$), see Fig.~\ref{som:fig:resonance_spectrum}(c). This indicates that the pair of atoms occupying the Laughlin state lives in a decoherence-free subspace, spanned by the states $|0\rangle  _\text{com} |\pm 2 \rangle _\text{rel}$, which is insensitive to environmental noise sources~\cite{Mamaev_2020, Hartke_2022}; here noise of the magnetic field.

The oscillation frequency of the Laughlin state $\delta \oR/2\pi = \SI{19.6(1)}{Hz} $ is roughly half the oscillation frequency of the $|\psi ^\text{(non-int)} \rangle$ state ${\delta \oR ^\text{(non-int)}/2\pi = \SI{42(1)}{Hz}}$ and of the ${|2\rangle _\text{com} |0\rangle _\text{rel}^\text{B}}$ state ${\delta \oR ^\text{(com)}/2\pi = \SI{42.0(2)}{Hz}}$. 
We argue that the effect can be qualitatively explained considering the detuning $\Delta'$ between the respective $2\hbar$ angular momentum eigenstate and closeby states with zero angular momentum, which determines the effective coupling between the $\ket{\pm}$ states (c.f. \cite{Lunt2024_pra}). In the case of the independent single-particle rotation at vanishing scattering length, the detuning is solely set by the anharmonicity in the trap shifting the resonant $\ket{n=2, m=0}$ state below the prepared $\ket{n=2, m=2}$ state. For the state ${|2\rangle _\text{com} |0\rangle _\text{rel}^\text{B}}$, all relevant states are interacting and hence shift together when the scattering length is tuned. Thus, we expect the relative detuning to change only slightly and $\delta \oR$ to stay approximately constant. In contrast, the Laughlin state is non-interacting (while states with zero relative angular momentum are interacting) which leads to a much larger detuning $\Delta'$ and in turn to a decrease of the oscillation frequency.

We measure the 2D densities of the Laughlin wavefunction during the time evolution in the slightly anisotropic optical potential for different quarter periods of the Ramsey delay time, shown in Fig.~\ref{som:fig:ramsey_spectrum}(f). The center-of-mass density remains stationary over time. The relative density evolves from the $\ket{2}$ state at $\tau = 0$, to an equal superposition of the $\ket{\pm 2}$ states at $\tau = T/4$, it continues to the $\ket{-2}$ state at $\tau = 2T/4$ and further evolves again to a superposition of the $\ket{\pm 2}$ states at $\tau = 3T/4$, however, now tilted by \SI{45}{\degree} with respect to the state at $\tau = T/4$.

\paragraph*{\textbf{Relative angle correlations}}

We calculate the normalized relative angle correlation function $g _{1/m} (\varphi)$ of the two-particle Laughlin wavefunction $\psi _{1/m}$ for arbitrary radii $r_\uparrow, r_\downarrow$ of the spin up and spin down atom, respectively. The unnormalized relative angle correlation function is defined as
\begin{align}
\begin{split}
    G_{1/m}(\varphi) & \equiv \int \text{d} r_\uparrow \int \text{d} \phi _\uparrow \int \text{d}r_\downarrow   \\
    & ~~~~ r_\uparrow r_\downarrow ~| \psi _{1/m} ( r_\uparrow , \phi _\uparrow , r_\downarrow , \phi _\uparrow - \varphi ) |^2 ,
\end{split}
\end{align}
where we integrate out the radial positions of the two particles $r_\uparrow, r_\downarrow$, as well as the absolute angle of the spin up fermion $\phi _\uparrow$. We then get 
\begin{align*}
    G_{1/m}(\varphi) & = 2 \pi  \int \text{d} r_\uparrow  \int \text{d}r_\downarrow   \\
    & ~~~~ r_\uparrow r_\downarrow ~ (  r_\uparrow ^2 + r_\downarrow ^2 - 2r_\uparrow  r_\downarrow  \cos \varphi  )  ^m e^{-(r_\uparrow ^2 + r_\downarrow ^2)} \\
    &= 2 \pi \sum _{k=0} ^{m} \sum _{q=0} ^{m-k} \binom{m}{k} \binom{m-k}{q} \int \text{d} r_\uparrow  \int \text{d}r_\downarrow ~  \\ 
    & ~~~~ r_\uparrow ^{2(m-k-q)+1} r_\downarrow ^{2q+1} (-2 r_\uparrow r_\downarrow \cos \varphi  )^k  e^{-(r_\uparrow ^2 + r_\downarrow ^2)} \\
    &= \frac{\pi}{2} \sum _{k=0} ^{m} \sum _{q=0} ^{m-k} \binom{m}{k} \binom{m-k}{q}  \\
    & ~~~~ (-2\cos \varphi )^k \Gamma \left(1 - \frac{k}{2} + m - q \right) \Gamma \left(1+\frac{k}{2}+q \right).
\end{align*}
We ensure that the correlation function $g _{1/m} (\varphi)$ is normalized accordingly 
\begin{align}
    g _{1/m} (\varphi) = \frac{ G_{1/m}(\varphi)  }{\int \text{d} \varphi ~ G_{1/m} (\varphi )}.
\end{align}

In our case of $m=2$, this yields the relative angle correlation function
\begin{align}
    g _ \text{1/2} (\varphi ) = \frac{6 - 3 \pi \cos( \varphi ) + 4 \cos ^2 (\varphi)}{16 \pi}.
\end{align}

\paragraph*{\textbf{Azimuthal distribution}}

\begin{figure}
    \centering
	\includegraphics{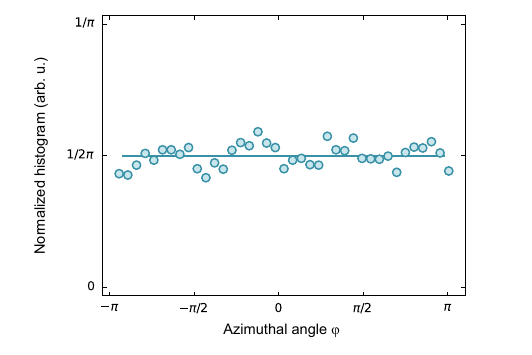}
    \caption{
    {Azimuthal distribution.} 
    Azimuthal distribution of the Laughlin state in the relative basis. We integrate over the radius $p_r$ and plot the normalized histogram over the azimuthal angles. The histogram is normalized to $1$. The horizontal line is at $1/2\pi$.
    \label{som:fig:azimuthal_inhomogenity}
    }
\end{figure}

We determine the azimuthal distribution of the Laughlin state in relative coordinates to estimate the contribution of the $|-2\rangle$ state, shown in Fig.~\ref{som:fig:azimuthal_inhomogenity}. In theory, the Laughlin state consists only of the $|2\rangle$ in relative coordinates, which is azimuthally homogeneous. A superposition of the $|\pm 2 \rangle$ states on the other hand shows oscillations in the density with respect to the azimuthal angle. We observe that the radially averaged histogram of the experimentally measured Laughlin wavefunction is approximately flat highlighting that it dominantly consists of the $|2\rangle$ state.

\paragraph*{\textbf{Molecule admixture}}

\begin{figure}
    \centering
	\includegraphics{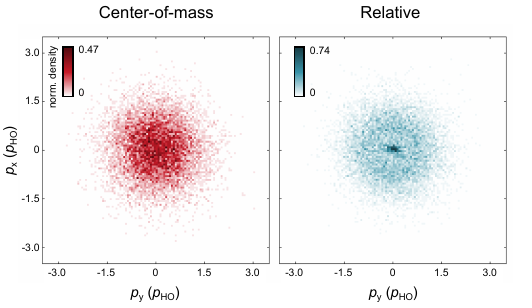}
    \caption{
    {Molecule admixture at 680G of the repulsively interacting ground state.} 
    Ramping the magnetic field from \SI{568}{G} (where the fermions are non-interacting) to repulsive interactions at \SI{680}{G} leads to coupling to molecular states with center-of-mass excitations. Measurement of the density of the wavefunction at \SI{680}{G} of the repulsively interacting ground state reveals the admixture of the Feshbach molecule -- a peak at zero momenta in the density in relative basis.
    \label{som:fig:molecule_680G_gs}
    }
\end{figure}

The anharmonicity of the optical potential couples the center-of-mass and relative motion which leads to coherent coupling of an unbound atom pair and a Feshbach molecule with center-of-mass excitation~\cite{Sala_PhysRevLett.110.203202}. The protocol for preparing the Laughlin state starts with an unbound atom pair in the ground state of the optical tweezer, at \SI{568}{G}, where the interaction strength vanishes. Subsequently, we ramp the magnetic field to \SI{680}{G} within \SI{10}{ms}. During this ramp we cross many molecular states with center-of-mass excitations. The anharmonicity leads to coherent coupling to these states.

In Fig.~\ref{som:fig:molecule_680G_gs}, we show the density distribution of the repulsively interacting ground state at $\SI{680}{G}$ in center-of-mass (red) and relative coordinates (blue). The Feshbach molecule is visible as a small peak at zero momenta in relative coordinates. 
The Laughlin state reported in the main text (c.f. Fig. 3) contains a $\sim \SI{1}{\%}$ admixture of such a Feshbach molecules. We conclude that the admixture of the Feshbach molecule is not a result of the rotating perturbation but of the magnetic field ramp.

\paragraph*{\textbf{Effective imaging resolution}}

We use the spin and atom resolved fluorescence imaging technique pioneered in previous work~\cite{Bergschneider_PhysRevA.97.063613}. In short, it consists of a time-of-flight expansion to magnify the initial wavefunction, followed by resonant imaging light the atoms absorb, re-emit and which is then detected on a camera to determine the initial momentum of each particle.

The imaging sequence starts after we prepared the desired angular momentum state in our cigar-shaped optical tweezer. We instantaneously switch off the optical tweezer and let the atoms expand. To compensate for the expansion in the axial direction, which would move the atoms out of the depth of focus of the objective, we ramp on an axial confinement (2D trap) with approximately the same axial trap frequency as the optical tweezer. This additional confinement is only turned on during the time-of-flight expansion. Hence, the atoms only expand radially in the remaining harmonic confinement of the 2D trap, with a radial trap frequency ${\omega _\text{2D,r} / 2 \pi  \approx \SI{85}{Hz}}$, and an axial trap frequency ${\omega _\text{2D,z} / 2 \pi  \approx \SI{5.6}{kHz}}$.

After the time-of-flight expansion, we use resonant light to illuminate the atoms for $\SI{15}{\mu s}$ in free-space without any pinning potential. We collect the fluorescence light ($\sim$20 photons) through an objective with $NA$=0.55 on an EMCCD camera allowing us to discriminate an atom from background noise. The two spin components are resolved by successively imaging the spin down fermion and the spin up fermion with a time delay of $\SI{145}{\mu s}$. From the camera we obtain the position of each spin state in units of pixels. The final size of the wavefunction is much larger than the initial size in the tweezer, therefore, we can map the final pixel position on the camera to the initial momentum of each atom, where we can compensate for the different expansion times of the spin states~\cite{Holten_2022}.

The in-situ imaging resolution is not sufficient to resolve the microscopic structure of the Laughlin wavefunction, since the harmonic oscillator length ${l_\text{HO} \approx \SI{0.17}{\mu m}}$, which is the natural length scale of the system, is much smaller than the imaging resolution ${\sigma _\text{l} \approx \SI{5.8}{\mu m}}$. 
Therefore, we let the system expand for $t_\text{tof} = \SI{1.78}{ms}$. The time-of-flight expansion maps the wavefunction in real space to momentum space; for harmonic oscillator states this yields a magnification ${M_\downarrow = \oR t_\text{tof} \approx \SI{623}{} }$ for the spin down atom and ${M _\uparrow = \oR t_\text{tof} \approx \SI{679}{} }$ for the spin up atom (due to the additional expansion time of $\SI{145}{\mu s}$). The effective imaging resolution in momentum space $\sigma _\text{p}$ in units of the harmonic oscillator momentum $p_\text{HO}$ is then given by ${\sigma _\text{p} /  p_\text{HO} = \sigma _\text{l} / l_\text{HO} ~ 1 / M}$.

\begin{figure}
    \centering
	\includegraphics{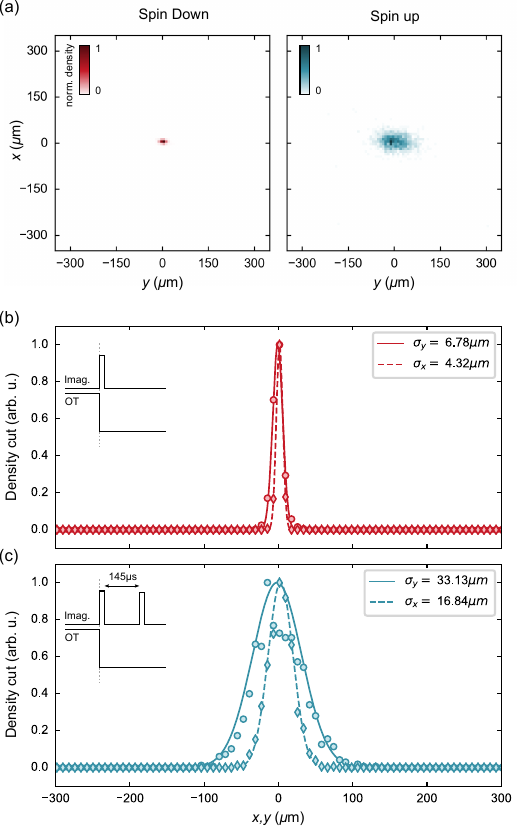}
    \caption{
    {Effective imaging resolution.}  
    (a) Normalized density distribution of the spin down (red) and spin up (blue) fermion with zero time of flight expansion. For better comparison, we show the same number of the camera pixels, as in all other figures in which density distributions are displayed; here expressed in the units of $\SI{}{\mu m}$.
    (b) Average imaging width of the spin down fermion. We find a small imaging anisotropy along the axis of the fluorescence imaging light.
    (c) Average imaging width of the spin up fermion. Here, the anisotropy is also visible. Additionally, off-resonant scattering during the imaging of the spin down fermion leads to broadening of the spin up fermion. 
    \label{som:fig:imaging_resolution}
    }
\end{figure}

In Fig.~\ref{som:fig:imaging_resolution}, we show the effective imaging resolution of the spin up and down fermion in units of $\SI{}{\mu m}$. We observe a small imaging anisotropy stemming from the direction of the imaging light. Furthermore, the imaging resolution of the spin up is broadened due to off-resonant scattering during the imaging of the spin down fermion. Finally, calculating the effective imaging resolution in units of the harmonic oscillator momentum of the spin down and spin up fermion results in ${2 \sigma ^\text{x} _{\text{p}\downarrow}  \approx \SI{0.08}{} p_\text{HO}}$, ${2   \sigma ^\text{y} _{\text{p}\downarrow}  \approx \SI{0.13}{} p_\text{HO}}$,  ${2 \sigma ^\text{x} _{\text{p}\uparrow}  \approx \SI{0.29}{} p_\text{HO}}$, ${2   \sigma ^\text{y} _{\text{p}\uparrow}  \approx \SI{0.57}{} p_\text{HO}}$.

\end{document}